\magnification=1200

\def\vspace#1{\crcr\noalign{\vskip#1\relax}}

\def\frac#1#2{{#1\over#2}}

\baselineskip=17 pt

\parskip=3pt plus 1pt

\leftskip=1.2 cm

\rightskip=20 truept

\vsize=8.6 true in

{\noindent\bf ELECTROMAGNETIC FIELD TENSOR AND LORENTZ FORCE AS CONSEQUENCE OF THE GEOMETRY OF
MINKOWSKIAN SPACETIME}

\bigskip

\centerline{J. Buitrago \footnote{${}^{\dag}$}{Permanent address: Departamento de
Astrof\'\i{}sica, Universidad de la La Laguna, 38200 La Laguna, Tenerife (Spain); e--mail:
jgb@ll.iac.es}  \& Chus Mart\'\i{}n \footnote{${}^{\ddag}$}{Area de F\'\i{}sica Te\'orica, Universidad de Salamanca, 37008
Salamanca (Spain); e--mail: chmm@gugu.usal.es}}

\noindent{\bf Abstract} We show that the electromagnetic field tensor and the Lorentz Force
are both a natural consequence of the geometric structure of Minkowskian space, being related
to infinitesimal boost and rotations in spacetime. The longstanding issue about the apparent
empirical origin of the Lorentz Force is clarified.

\bigskip

\centerline{ Submitted to {\it Journal of Mathematical Physics\/}}

\bigskip

Jes\'us Buitrago

Departamento de Astrof\'\i{}sica

Universidad de La Laguna

38200 La Laguna, Tenerife (Spain)

e--mail: jgb@ll.iac.es

\vfill\eject

{\noindent\bf I. INTRODUCTION}

\bigskip

It is a well known fact that the electromagnetic force acting on a charged
particle cannot be obtained from Maxwell Equations  having, ultimately, an
empirical basis (at least from a classical standpoint). This problem have been
around for many years, leaving the electromagnetic field (which we have always
considered our better understood part of Physics) with an apparent, and rather
curious, lack of self-consistency, specially if we compare the situation with
its equivalent in General Relativity where the Equation of Geodesics follows
both from the underlying Riemannian geometry as well as from Einstein
Equations. 

During the last decades, the subject seems to have been put aside, as a peculiar
feature of the theory rather than a problem. However, its real magnitude comes
out when we consider the efforts of Einstein and Weyl (among others), some 80
years ago, of unifying gravitation and electromagnetism in a common geometrical
framework. Although one can argue that Einstein and Weyl attempts were far more
ambitious than simply obtaining the Lorentz Force, it is also true that Weyl's
essential result was the derivation of the electromagnetic field tensor on a
geometric basis (We note that his theory, apart of leading to unphysical predictions, also
failed to obtain the Lorentz Force)(See [1] and references therein to Weyl's original work).

It is the aim of this article to show that the electromagnetic field tensor and
the Lorentz Force are a consequence of the geometric structure of Minkowskian
space having its origin in infinitesimal spatial rotations and boosts in
spacetime. The Lorentz Force is thus also susceptible of a geometrical
interpretation and the electromagnetic theory is self-consistent. A heuristic
approach to the problem was already published by one of us a few years ago [2].
The scope of this article is to present a rigorous derivation.

\bigskip

{\noindent\bf II. INTEGRAL CURVES ASSOCIATED WITH THE EQUATIONS OF MOTION AND
FINITE LORENTZ TRANSFORMATIONS}

\bigskip

In the Minkowskian spacetime $M_4$, the equations of motion related to a
certain particle following an arbitrary trajectory form a second order
differential system:
$$
\left\{\eqalign{
dx^\alpha\over d \tau &= u^\alpha \cr
\vspace{2mm} 
du^ \alpha\over d\tau &= h^\alpha (x,u)
}\right.
\eqno(1)$$
where $x^\alpha$ is the position of the particle, $u^\alpha$ is  the
four-velocity and $\tau$ the proper time.

If $x_0$ is a generic point in spacetime, we write the general integral of (1)
as
$$
\left\{\eqalign{
x^\alpha=f^\alpha (x_0, u_0;\tau)\cr
\vspace{2mm}
u^\alpha=\dot f^\alpha(x_0,u_0;\tau)\cr
}\right.
\eqno(2)$$
($\cdot \equiv {d\over d\tau}$) such that $f^\alpha(x_0,u_0;0)
\equiv x_0^\alpha$, $\dot f^\alpha(x_0,u_0;0)\equiv u_0^\alpha$.

\bigskip

We thus regard the solutions of (1) as parametric curves in $M_4$. As the
diferential system (1) is autonomous, the integral curves (2) generate a
uniparametric group [3]:
$$
\left\{\eqalign{
f^\alpha[f(x_0,u_0;\tau_1),\dot f(x_0,u_0;
\tau_1);\tau_2]=f^\alpha(x_0,u_0;\tau_1+\tau_2)\cr
\vspace{2mm}
\dot f^\alpha[f(x_0,u_0;\tau_1),\dot f(x_0,u_0;
\tau_1);\tau_2]=\dot f^\alpha(x_0,u_0;\tau_1+\tau_2)\cr
}\right.
\eqno(3)$$
for any $\tau_1, \tau_2$.

\bigskip

Given a spacetime trajectory $x^\alpha=f^\alpha(x_0,u_0;\tau)$ 
travelled by the particle with a tangent four-velocity vector
$u^\alpha(\tau)=\dot f^\alpha(x_0,u_0;\tau)$ and adopting a matrix
representation of the homogeneous Lorentz Group, we can write:
$$
\dot f^\alpha(x_0, u_0;\tau)=L^\alpha_\lambda(x_0;\tau)u_0^\lambda,
\eqno(4)
$$
$L^\alpha_\lambda$ being a matrix of the  Lorentz Group  which transport the
four velocity vector from a point $x_0$, where the four-velocity is
$u^\lambda_0$, to another  point $x$, on the curve
$x^\alpha=f^\alpha(x_0,u_0;\tau)$, where it is $u^\lambda(\tau)$. (Note that
we adopt the "active" point of view and consider $L^\alpha_\lambda$ as an
operator acting on $u^\lambda$). 
Furthermore, as it is  readily seen from the second relation in (3) and the last equation,
the group condition (3) implies that the 
$\tau$--parametric family of Lorentz matrices $L^\alpha_\lambda(x_0;\tau)$
verify the relation  
$$
L(\tau_1)\cdot L(\tau_2) = L(\tau_1 + \tau_2),
$$
meaning that this family is an uniparametric subgroup of the Lorentz Group.

\bigskip

In principle, $L^\alpha_\lambda$ could be any (or any product of them) of the
well known matrices which, for completeness, we list below :
$$
{L_{(1)}}^\alpha_\lambda=\pmatrix{1&0&0&0\cr
0&1&0&0\cr
0&0&\cos\varphi_1(x_0,\tau)&-\sin\varphi_1(x_0,\tau)\cr
0&0&\sin\varphi_1(x_0,\tau)&\cos\varphi_1(x_0,\tau)\cr} \eqno(5)
$$

\bigskip

$$
{L_{(2)}}^\alpha_\lambda=\pmatrix{1&0&0&0\cr
0&\cos\varphi_2(x_0,\tau)&0&\sin\varphi_2(x_0,\tau)\cr
0&0&1&0\cr
0&-\sin\varphi_2(x_0,\tau)&0&\cos\varphi_2(x_0,\tau)\cr} \eqno(6)
$$

\bigskip 

$$
{L_{(3)}}^\alpha_\lambda=\pmatrix{1&0&0&0\cr
0&\cos\varphi_3(x_0,\tau)&-\sin\varphi_3(x_0,\tau)&0\cr
0&0&1&0\cr
0&\sin\varphi_3(x_0,\tau)& \cos\varphi_3(x_0,\tau)&0\cr
0&0&0&1\cr} \eqno(7)
$$

\bigskip

$$
{L_{(4)}}^\alpha_\lambda=\pmatrix{\cosh \psi_1(x_o,\tau)&\sinh
\psi_1(x_0,\tau)&0&0\cr
\sinh\psi_1(x_0,\tau)&\cosh \psi_1(x_o,\tau)&0&0\cr
0&0&1&0\cr
0&0&0&1\cr} \eqno(8)
$$

\bigskip

$$
{L_{(5)}}^\alpha_\lambda=\pmatrix{\cosh \psi_2(x_o,\tau)&0&
\sinh\psi_2(x_0,\tau)&0\cr
0&1&0&0\cr
\sinh\psi_2(x_0,\tau)&0&\cosh \psi_2(x_0,\tau)&0\cr
0&0&0&1\cr} \eqno(9)
$$

\bigskip

$$
{L_{(6)}}^\alpha_\lambda=\pmatrix{\cosh \psi_3(x_0,\tau)&0&0&
\sinh\psi_3(x_0,\tau)\cr
0&1&0&0\cr
0&0&1&0\cr
\sinh\psi_3(x_0,\tau)&0&0&\cosh \psi_3(x_0,\tau)\cr} \eqno(10)
$$

\bigskip

${L_{(i)}}^\alpha_\lambda, (i=1,2,3)$ describe the 3 finite spatial rotations
around the $Ox^1$, $Ox^2$, $Ox^3$ axis while the other three remaining are associated
to Lorentz boosts along the same axis. $\varphi_i(x_0,\tau)$ and $\psi_i(x_0,\tau)$
being the turning angles with their corresponding arguments.

\bigskip

{\noindent\bf III. INFINITESIMAL TRANSFORMATIONS, MAXWELL TENSOR AND THE
ELECTROMAGNETIC FORCE}

\bigskip

As already stated in the introduction, in this section we shall show how 
infinitesimal transformations in spacetime can be related  to our familiar
notions of electric and magnetic fields. 

We first notice that since the rotation of the spatial part of
$u^\lambda(\tau)$, between $x_0$ and $x$, can proceed at an arbitrary rate, the
rotation angle $\varphi_i(x_0,\tau)$ appearing in the first three matrices
above can be written as
$$
\varphi_i(x_0,\tau)=\bar b_i(x_0)\cdot\tau  \quad,\quad (i=1,2,3)\eqno(11)
$$
$\bar b_i(x_0)$ meaning the average rotation (around the $Ox_i$ axis) per
unit time during the lapse of proper time $\tau$.

In the same way, for the boost "angle" in the other matrices we make
$$
\psi_i(x_0,\tau)=\bar \epsilon_i(x_0)\cdot\tau, \eqno(12)
$$
with the same interpretation but this time in terms of space-time "rotations".

We now note that the more general  lorentz transformation (which we call
${L_G}^\alpha_\lambda$) that can act on the four velocity vector is given as
the product
$$
{L_G}(x_0;\tau)=\prod_{i=1}^6 {L_{(i)}}(x_0;\tau). \eqno(13)
$$
then, using the preceding parametrization, we can substitute (4) by:
$$
\dot f^\alpha(x_0;\tau)={L_G}^\alpha_\lambda(x_0;\tau)
u_0^\lambda, \eqno(14)
$$    

Now as ${L_G}(\tau)$ is an uniparametric subgroup of the Lorentz Group,
according to the theory of Lie groups, its derivative, at $\tau=0$, is an element of its Lie
Algebra [3]. Although quite a tedious calculation involving multiplication of
the above six matrices (after inserting expressions (11) and (12)), derivation
respect to
$\tau$ and going to the limit $\tau=0$, it is possible to arrive at the
following result:
$$
{d {L_G}^\alpha_\lambda \over d\tau}\vert_{\tau=0}=
\pmatrix{0&\epsilon_1(x_0)&\epsilon_2(x_0)&\epsilon_3(x_0)\cr
\epsilon_1(x_0)&0&-b_3(x_0)&b_2(x_0)\cr
\epsilon_2(x_0)&b_3(x_0)&0&-b_1(x_0)\cr
\epsilon_3(x_0)&-b_2(x_0)&b_1(x_0)&0\cr} \equiv Q^\alpha_\lambda(x_0)\eqno(15)
$$
where $b_i(x_0)$ and $\epsilon_i(x_0), (i=1,2,3)$ are now the infinitesimal
spatial rotation rate and boost rate at $x_0$.

From Eqs (1), (2) and (14): 
$$
\dot u_0^\alpha= h^\alpha(x_0,u_0)=\ddot f^\alpha(x_0;\tau). \eqno(16)
$$
Therefore, we get (since $x_0$ is a  generic point)
$$
{du^\alpha\over d\tau}=Q^\alpha_\lambda \cdot u^\lambda \quad, \eqno(17)
$$

This last equation tell us how the structure of Minkowskian spacetime 
{\it prescribe a linear evolution} of  the four-velocity vector. Since the
evolution should actually take place as consequence of the action of an external
field, then, up to a proportionality constant, the "rotation" rate $\epsilon_i$
can be identified with the {\it i}-component of     the electric field and
the spatial rotation rate $b_i$ with the {\it i}-component of the magnetic field:
$$
\epsilon_i=k E_i
$$
$$
b_i=k B_i
$$
so that instead of (17) we can write 
$$
{du^\alpha\over d \tau}=k F^\alpha_\lambda u^\lambda, \eqno(18)
$$
with $F^\alpha_\lambda$ the Maxwell field tensor which by virtue of its
definition, via Eq. (15),  must comply  with
$$F^\alpha_\lambda \eta_{\alpha\mu}+ F^\alpha_\mu \eta_{\alpha\lambda}=0,
\eqno(19)
$$
merely a statement of the antisymmetry of $F_{\lambda\mu}$.

Finally, as $F_{\alpha\mu}$ is an element of the Lie Algebra, under a finite Lorentz boost,
transform by means of the self-adjoint representation [3], thus giving the right rule of
change for every component of both $\vec E$ and $\vec B$ between inertial frames.

In this new picture of the electromagnetic field, $\vec E$ and $\vec B$ should
be regarded, respectively, as performers of infinitesimal boosts and spatial rotations in
spacetime. The new image is a manifestation of the deep harmony between special relativity
and electromagnetism.

ACKNOWLEDGEMENTS

J. B. wants to express his gratitude to the Group of Theoretical Physics of the
University of Salamanca for his hospitality during the elaboration of this work, which was
supported by a grant (PB96--1306) from ``Ministerio de Educaci\'on y Cultura" of Spain. 

\vfill\eject

{\noindent\bf REFERENCES}

[1]W. Pauli, {\it Theory of Relativity\/}. Dover Publications, Inc., New York (1958)

[2]J. Buitrago, {\it Electromagnetic force and geometry of Minkows\-kian spa\-ce--time\/},
European Journal of Physics {\bf 16}, 113 (1995)

[3] Choquet--Bruhat, Y (1968). {\it G\'eom\'etrie diff\'erentielle et syst\`emes
exterieurs\/} (Dunod, Paris).

\end